\documentclass[12pt]{article}
\usepackage{amsmath, amssymb, slashed, epsf, color}
\usepackage{geometry}
\usepackage{fullpage}
\usepackage{mathabx}
\usepackage[title]{appendix}

\ifpdf
	\usepackage{graphicx}
	\usepackage[colorlinks=true]{hyperref}
\else
	\usepackage[dvipdfmx]{graphicx}
	\usepackage[dvipdfmx, colorlinks=true]{hyperref}
\fi

\def\beq{\begin{equation}}
\def\eeq{\end{equation}}
\def\mrm{\mathrm}

\begin{document}

\begin{titlepage}
\begin{center}
\hfill CPHT-RR034.052018 \\
\hfill UMN--TH--3718/18, FTPI--MINN--18/08 \\ 
\hfill  LPT-Orsay-18-62

\vspace{1.0cm}
{\Large\bf Gravitino Decay in High Scale Supersymmetry \\with R-parity Violation}

\vspace{1.0cm}
Emilian Dudas$^{a}$,
Tony Gherghetta$^{b}$,
Kunio Kaneta$^{b,c}$,\\
Yann Mambrini$^{d}$,
and
Keith A. Olive$^{b,c}$

\vspace{0.5cm}
{\it
${}^a$CPhT, Ecole Polytechnique, 91128 Palaiseau Cedex, France \\
${}^b$School of Physics and Astronomy, University of Minnesota, Minneapolis, MN 55455, USA\\
${}^c$William I.~Fine Theoretical Physics Institute, University of Minnesota, Minneapolis, MN 55455, USA\\
${}^d$Laboratoire de Physique Th\'eorique 
Universit\'e Paris-Sud, F-91405 Orsay, France
}

\vspace{0.5cm}
\abstract{
We consider the effects of R-parity violation due to the inclusion of a bilinear $\mu^\prime L H_u$ superpotential term in high scale supersymmetric models with an EeV scale gravitino as dark matter. Although the typical phenomenological limits on this coupling (e.g. due to lepton number violation and the preservation of the baryon asymmetry) are relaxed when the supersymmetric mass spectrum is assumed to be heavy 
(in excess of the inflationary scale of $3 \times 10^{13}$ GeV), the requirement that the gravitino be sufficiently long-lived so as to account for the observed dark matter density, leads to a relatively strong bound on $\mu^\prime \lesssim 20$ GeV.
The dominant decay channels for the longitudinal component of the gravitino are 
$Z \nu, W^\pm l^\mp$, and $h\nu$. To avoid an excess neutrino signal in IceCube, our limit on $\mu'$ is then strengthened to $\mu' \lesssim 50$ keV. When the bound is saturated, we find that there is a potentially detectable flux of mono-chromatic neutrinos with EeV energies.
 }

\end{center}
\end{titlepage}
\setcounter{footnote}{0}

\section{Introduction}
\label{sec:introduction}

Naturalness and potential solutions to low energy phenomenological quandaries such 
as the discrepancy between the theoretical and experimental determinations of the 
anomalous magnetic moment of the muon \cite{newBNL,g-2} pointed to low energy
supersymmetry. Indeed, statistical analyses of a multitude of low energy observables 
predicted \cite{mc3,fittino} a supersymmetric spectrum well within reach of the LHC.
However, to date, there has been no experimental confirmation of low energy
supersymmetry \cite{nosusy}. Supersymmetry may still lie within the reach
of the LHC, and discovery may occur in upcoming runs. Nevertheless it is also possible that supersymmetry lies beyond the LHC reach, and in that case, 
it is unclear whether the scale of supersymmetry, ${\widetilde m}$ is just beyond its reach, ${\widetilde m} \sim 10$ TeV, or far beyond its reach, ${\widetilde m} > 10^{13}$ GeV, for example. 

If supersymmetry plays a role in nature below the Planck scale, it may
still be broken at some high energy scale \cite{hssusy}. 
If that scale is above the inflationary scale, $\sim 3 \times 10^{13}$ GeV,
supersymmetric particles, with one exception, may not have participated in the reheating process
and were never part of the thermal background in the early Universe. The exception may be
the gravitino with an approximately EeV mass\,\cite{dmo}. In this case, we still have a viable supersymmetric
dark matter candidate, namely the gravitino which {\em is} produced from the thermal bath during reheating \cite{bcdm,dmo,dgmo}.
Interestingly, in the context of an SO(10) GUT, such high scale supersymmetric models are still
able to account for gauge coupling unification, radiative electroweak symmetry breaking and the
stability of the Higgs vacuum \cite{egko}.
However if the gravitino is stable, as would be the case if $R$-parity is conserved, 
there are very few detectable signatures of the model.

$R$-parity is typically imposed in supersymmetric models to insure the stability of the proton 
\cite{Farrar:1978xj} by eliminating all baryon and lepton number violating operators. 
Of course, a consequence of $R$-parity conservation is  that the lightest supersymmetric particle (LSP) is stable and
becomes a dark matter candidate \cite{ehnos}. 
Limits on $R$-parity violating (RPV) couplings can be derived by requiring baryon
and lepton number violating interactions to remain out-of-equilibrium
in the early universe to preserve the baryon asymmetry \cite{cdeo}.
However in high scale supersymmetry, these limits are relaxed as supersymmetric partners
were never in the thermal bath and did not mediate interactions which could 
wash out the baryon asymmetry. Therefore, it is possible that some
amount of RPV is acceptable.
If present, RPV violating operators would render the lightest
supersymmetric particle, the gravitino in this case, unstable.
If long-lived, the decay products may provide a signature for the EeV gravitino.

In this paper we consider a minimal addition to the minimal supersymmetric standard model (MSSM).
Namely, we include a single RPV interaction, generated by 
the $LH_u$ bilinear term in the superpotential. This term is sufficient to allow for the decay of the LSP gravitino, and demanding that it remains long-lived to account 
for the dark matter,
will enable us to set a limit on the ``$\mu$"-term associated with this bilinear. We will compare this limit with the one imposed
from the preservation of the baryon asymmetry in both weak scale and high scale supersymmetry models. Furthermore, 
as we will show, while there is a $\gamma\nu$ decay mode, the dominant 
decay channel actually proceeds through the longitudinal mode of the gravitino
to $Z \nu$, $W^\pm l^\mp$, and $h\nu$. Thus this model predicts a monochromatic source of $\sim$ EeV neutrinos.

The paper is organized as follows. In the next section
we discuss the expected abundance of the heavy gravitino
produced during reheating. We also make some preliminary remarks 
concerning the expected effects of including the $LH_u$ RPV term.
In section 3, we introduce the $LH_u$ term and discuss its role in the neutralino and chargino mixing matrices and its role as a 
source for neutrino masses. Constraints arising from other relevant operators are also discussed. In section 4, we compute the lifetime and branching ratios 
of the gravitino and in section 5, we discuss the observational consequences of its decay.
Our conclusions are summarized in section 6.

\section{Some Preliminaries}

Generically, in weak scale supersymmetry models with a gravitino LSP,
the gravitino mass is typically $\mathcal{O}(100)$ GeV.
Higher masses lead to an overabundance of gravitinos,
independent of the reheat temperature due to the decays 
of the next-to-lightest supersymmetric particle, often a neutralino, to the gravitino.
It is difficult to obtain neutralinos with masses in excess of a few TeV, with relic densities still compatible with CMB observations \cite{eeloz}.
By combining the limit on the relic density with limits from big bang nucleosynthesis, one can 
derive an upper limit of roughly 4 TeV on the gravitino mass \cite{dmo}. This limit is evaded in high scale supersymmetry models,
when no superpartners other than the gravitino are produced
during reheating and a new window of gravitino masses opens up
above $\mathcal{O}(0.1)$ EeV \cite{dmo}.

In high scale supersymmetry models with the gravitino as the only 
superpartner lighter than the inflaton, gravitinos 
can be pair produced during reheating \cite{bcdm,dmo}.
The gravitino production rate density was derived in 
\cite{bcdm}
\beq
R = n^2 \langle \sigma v \rangle \simeq 2.4 \times \frac{T^{12}}{M_P^4 m_{3/2}^4}\,,
\label{Eq:r}
\eeq
where $M_P = 2.4 \times 10^{18}$ GeV is the reduced Planck mass, and $n$ is the number density of incoming states.
The gravitino abundance can be determined by comparing the 
rate $\Gamma \sim R/n \sim T^9/M_P^4 m_{3/2}^4$ to the Hubble expansion rate so that $n_{3/2}/n_\gamma \sim \Gamma/H \sim T^7/M_P^3 m_{3/2}^4$.
More precisely, we find,
\beq
\Omega_{3/2}h^2 
\simeq
0.11 \left( \frac{0.1 ~\mathrm{EeV}}{m_{3/2}} \right)^3
\left( \frac{T_{RH}}{2.0 \times 10^{10}~\mathrm{GeV}} \right)^7 \, .
\label{Eq:omega}
\eeq
In the absence of direct inflaton decays, we see that a 
reheating temperature, $T_{RH}$, of roughly $10^{10}$ GeV is required.
This was shown to be quite reasonable in a more detailed
model which combined inflation with supersymmetry breaking \cite{dgmo}. In that model, the dominant mechanism for reheating 
involved inflaton decays to Standard Model Higgs pairs. 

Without R-parity violation, the gravitino remains stable and
experimental signatures are limited.  Instead R-parity violation
allows the possibility for gravitino decays and
perhaps an indirect signature for gravitino dark matter.
Here, we concentrate on the effects of adding an $LH_u$ 
term to the superpotential leading to decays such as
$\nu \gamma$, $\nu Z$, $\nu h$, and $l W$.
We next outline the channels we expect to dominate in gravitino decay.
Our argument here will be largely heuristic and a more detailed
derivation follows in section 4.

To estimate the decay width, one can consider the coupling of the gravitino, $\psi_\mu$ to a massive gauge field. For simplicity, we consider the abelian Higgs model with a $U(1)$ gauge group. The coupling is generated through the gravitational interaction
\begin{equation}
{\cal L }_{int} = \frac{-i}{\sqrt{2} M_P}D_\mu \phi^\dagger \,{\bar\psi}_\nu \gamma^\mu \gamma^\nu\,\chi_{L} + h.c.\, ,
\end{equation}
between the gravitino, $\psi_\nu$, the Higgs field, $\phi$ and Weyl fermion, $\chi_{L}$ (which plays the role of the Higgsino). The Lagrangian can be written as function of the Goldstino component, $\psi$ of the gravitino, and the Higgs field components
\beq
\psi_\nu \sim \frac{\partial_\nu \psi}{m_{3/2}}\quad {\rm or}\quad i\gamma_\nu \psi; 
~~~\phi = \frac{1}{\sqrt{2}} (v+h) e^{-i \frac{\theta}{v}}
\, ,
\eeq
where $v$ is the Higgs vacuum expectation value, $h$ is the radial component (Higgs boson) and $\theta$ the corresponding Nambu-Goldstone boson. A more detailed calculation (see the Appendix) shows that the dominant contribution arises from $\gamma_\nu \psi$, leading to the interaction
\beq
{\cal L}_{int} \sim \frac{1}{M_P} \partial_\mu\theta \,{\bar\psi} \gamma^\mu\chi_L+ h.c.\,.
\eeq
In the massless $\chi_{L}$ limit, the amplitude squared then becomes\footnote{As will be shown in the Appendix, the piece $\psi_\nu\sim\partial_\nu\psi/m_{3/2}$ leads to $|{\cal M}|^2 \sim m_{3/2}^2 m_A^2/M_P^2$ where $m_A$ is the gauge boson mass, which is highly suppressed when $m_A\ll m_{3/2}$.}
\beq
|{\cal M}|^2 \sim \frac{m_{3/2}^4}{M_P^2} \, . 
\eeq
Anticipating that the $LH_u$ term will induce a mixing, parameterized by $\epsilon$, between $\chi_L$ (or the Higgsino) and the neutrino (to be discussed in detail below), we can write $\chi_{L}\sim \epsilon\,\nu$. The dominant decay channel is then $\psi_\mu \rightarrow \nu Z/h$, with a width
\beq
\Gamma_{3/2} \sim \frac{|{\cal M}|^2}{s}m_{3/2} \sim \epsilon^2 \frac{m_{3/2}^3}{M_P^2 } \, .
\label{main} 
\eeq
From the above argument, we can also anticipate that the Goldstino decay to $\nu \gamma$ will be suppressed since the photon does not have a longitudinal component. In the detailed calculation the result (\ref{main}) will be generalized to the non-Abelian, supersymmetric two Higgs doublet case. 
In section \ref{OC}, we will derive limits on $\epsilon$ 
from existing experimental constraints, requiring in addition,
that sufficiently many gravitinos are present today to supply the 
dark matter. 

\section{R-Parity Violation}
\label{sec:r_parity_violation}

The simplest model including RPV only introduces a bilinear RPV operator:
\begin{eqnarray}
	W &=& W_{\rm MSSM} + W_{\rm RPV}, \\
	W_{\rm MSSM} &=& \mu H_u H_d + y_e L H_d e^c + y_u Q H_u u^c + y_d Q H_d d^c, \label{Wmssm}\\
	W_{\rm RPV} &=& \mu' L H_u \label{Wrpv}.
\end{eqnarray}
In general the RPV mass parameter $\mu'$ depends on the lepton flavor, but here we omit the flavor dependence for simplicity (for more detailed discussion, see, e.g., \cite{Barbier:2004ez}). Note that we have suppressed all generation indices in both (\ref{Wmssm}) and (\ref{Wrpv}).
Since lepton number is no longer conserved, $L$ and $H_d$ cannot be distinguished in this setup, and thus there is a field basis dependence in defining the $L$ and $H_d$ fields.
For instance, when we take $L \to c_\xi L + s_\xi H_d$ and $H_d \to c_\xi H_d - s_\xi L$ with $s_\xi=\sin\xi$, $c_\xi=\cos\xi$ and $\tan\xi=\mu'/\mu$, we can eliminate the bilinear RPV term.
Instead, we obtain trilinear RPV terms, such as $y_e s_\xi LLe^c$ and $y_d s_\xi QLd^c$.
Though the observables do not depend on our choice of basis, we need to clarify which basis we use.
We will work in a basis where we define the linear combination of the four fields, $L$ and $H_d$, which picks up a vacuum expectation value to be the Higgs
and write Eq. (\ref{Wrpv}) without any additional trilinear terms. 
In either case, while lepton number is violated, baryon number is still conserved, so this model is free from proton decay constraints.
In the following calculation, we will take the basis that explicitly keeps only the bilinear term given in $W_{\rm RPV}$.

\subsection{Induced Mixing}
\label{sec:inducedmixing}

The inclusion of the RPV bilinear term induces a mixing between the charged leptons and the charged Higgsinos. 
In the relevant fermionic part of the Lagrangian, the mass matrix for the charged fermions in the form of ${\cal L}_{\rm mass} = -(\tilde W^+, \widetilde H_u^+, l^c) {\cal M}_C (\tilde W^-, \widetilde H_d^-, l)^T + h.c.$ is given by
\begin{eqnarray}
	{\cal M}_C =
	\left(
		\begin{array}{ccc}
			M_2 & g v_d & 0 \\
			g v_u & \mu &  \mu'  \\
			0 &0 & y_e v_d
		\end{array}
	\right).
    \label{chargino}
\end{eqnarray}
Without loss of generality, we can take a lepton field basis such that $y_e$ becomes diagonal.
For the neutral fermions, the mass matrix in the field basis $(\tilde B, \tilde W^0, \widetilde H_u^0, \widetilde H_d^0, \nu)$ is given by
\begin{eqnarray}
	{\cal M}_N &=&
	\left(
		\begin{array}{ccccc}
			M_1 & 0& \frac{g' v_u}{\sqrt2} & -\frac{g' v_d}{\sqrt2} &0\\
			0& M_2 & -\frac{g v_u}{\sqrt2} & \frac{g v_d}{\sqrt2} &0\\
			\frac{g' v_u}{\sqrt2} & -\frac{g v_u}{\sqrt2} & 0& -\mu & -\mu'\\
			-\frac{g' v_d}{\sqrt2} & \frac{g v_d}{\sqrt2} & -\mu & 0&0 \\
			0& 0& -\mu' &0 &0
		\end{array}
	\right)
	\equiv
	\left(
		\begin{array}{cc}
			\hat M & \hat m\\
			\hat m^T & \hat\mu
		\end{array}
	\right),	
\end{eqnarray}
where we have defined
\begin{eqnarray}
	\hat M =
	\left(
		\begin{array}{cc}
			M_1 & 0\\
			0 & M_2
		\end{array}
	\right),~
	\hat m =
	\left(
		\begin{array}{ccc}
			\frac{g' v_u}{\sqrt2} & -\frac{g' v_d}{\sqrt2} & 0 \\
			-\frac{g v_u}{\sqrt2} & \frac{g v_d}{\sqrt2} & 0
		\end{array}
	\right),~
	\hat \mu =
	\left(
		\begin{array}{ccc}
			0 & -\mu & -\mu'\\
			-\mu & 0 & 0\\
			-\mu' & 0 & 0
		\end{array}
	\right).
\end{eqnarray}
Now it is clear that the neutrino acquires a mass due to a non-vanishing $\mu'$, which is given by
\beq
m_\nu \simeq \epsilon^2 c_\beta^2 
\left(\frac{c_W^2}{M_2}+\frac{s_W^2}{M_1} \right) M_Z^2\,,
\label{numass}
\eeq
where $c_\beta=\cos\beta$ with $\tan\beta = v_u/v_d$, $\tan \theta_W = g'/g$, $s_W=\sin\theta_W$, $c_W=\cos\theta_W$, $\epsilon = s_\xi =\mu'/{\bar \mu}\equiv \mu'/\sqrt{\mu^2 + \mu'^2} \approx \mu'/\mu$ when $\mu' \ll \mu$ as will assume later. Note that this mass is too small to account for the physical neutrino masses.
To derive Eq. (\ref{numass}) we diagonalized the mass matrix perturbatively as follows:
suppose a unitary matrix $U$ diagonalizes ${\cal M}_N$ as $U^T {\cal M}_NU = {\cal M}_N^{\rm diag}$. We may take
\begin{eqnarray}
	U &=& \left(
		\begin{array}{cc}
			1_{2\times2} & 0\\
			0 & V
		\end{array}
	\right)
    \exp
    \left(
		\begin{array}{cc}
			0 & \theta \\
			-\theta^T & 0
		\end{array}
	\right)
    \simeq
	\left(
		\begin{array}{cc}
			1_{2\times2} & 0\\
			0 & V
		\end{array}
	\right)
	\left(
		\begin{array}{cc}
			1-\frac{1}{2}\theta\theta^T & \theta \\
			-\theta^T & 1-\frac{1}{2}\theta^T\theta
		\end{array}
	\right),
\end{eqnarray}
where $\theta$ and $V$ are $2\times 3$ and $3\times 3$ matrices, respectively. The matrix $V$ satisfies $V^T \hat\mu V = \hat\mu^{\rm diag}$, which allows $V$ to be written as a function of $\mu$ and $\mu'$, given by\footnote{We have taken the neutrino component as a massless eigenstate.} 
\begin{eqnarray}
		V = \frac{1}{\sqrt 2}
	\left(
		\begin{array}{ccc}
			1 & -1 & 0\\
			c_\xi & c_\xi & -\sqrt{2}s_\xi\\
			s_\xi & s_\xi & \sqrt{2}c_\xi
		\end{array}
	\right) \, .
\end{eqnarray}

The matrix $\theta$ can be obtained by solving the conditions $[U^T {\cal M}_N U]_{ij} = 0$, $i\ne j$.\footnote{In the following calculation we neglect ${\cal O}(\theta^2)$ terms in solving $[U^T {\cal M}_N U]_{ij}=0$ to get $\theta$. Indeed, this is a good approximation as long as $\widetilde m$ is sufficiently large compared to the weak scale.}
In this parametrization the solution is 
\begin{eqnarray}
	\theta &=&
    \frac{1}{\sqrt 2}
	\left(
		\begin{array}{ccc}
			\frac{M_Z}{M_1+\bar\mu}s_W(c_\beta c_\xi-s_\beta) & \frac{M_Z}{M_1-\bar\mu}s_W(c_\beta c_\xi+s_\beta) & -\frac{\sqrt{2}M_Z}{M_1}s_W c_\beta s_\xi\\
            -\frac{M_Z}{M_2+\bar\mu}c_W(c_\beta c_\xi-s_\beta) & -\frac{M_Z}{M_2-\bar\mu}c_W(c_\beta c_\xi+s_\beta) & \frac{\sqrt{2}M_Z}{M_2}c_W c_\beta s_\xi
		\end{array}
	\right),
    \label{theta}
\end{eqnarray}
where $s_\beta= \sin\beta$.
Then, by ignoring the mixing with gauginos, namely, at the leading order in the perturbative diagonalization, the mass eigenstate $\chi_0 \equiv (\tilde h, \widetilde H, \nu)$ is related to the gauge eigenstate $N \equiv (\widetilde H_u^0, \widetilde H_d^0, \hat\nu)$ as $\chi_0 = V^T N$, so, for instance, $\nu$ is given by $\nu = (-s_\xi \widetilde H_d^0 + c_\xi \hat\nu)/\sqrt{2}$ and contains no $\widetilde H_u^0$ component.
Similarly, it is clear that $\widetilde H_u^0$ does not have a $\nu$ component to order $\theta$, thus the non-zero contribution in $U_{\widetilde H_u^0\nu}$ comes from the perturbation in $\theta^T\theta = {\cal O}(M_Z^2/{\widetilde m}^2)$. In contrast, the $\widetilde H_d^0$ has a term $-s_\xi \nu$ which is the leading contribution in $U_{\widetilde H_d^0\nu}$.
Therefore, at leading order, $U_{\widetilde H_u^0\nu}$ is suppressed by a factor of $(M_Z/{\widetilde m})^2$ compared to $U_{\widetilde H_d^0\nu}\sim s_\xi\,(=\epsilon)$.

Since we are focusing on the case that all sparticles except for the gravitino are heavier than the inflation scale ($\sim 3\times 10^{13}$ GeV), the constraint on the neutrino masses, $\sum_i m_{\nu_i} < 0.151$ eV (95\% CL) \cite{vagnozzi} or similar limits \cite{Giusarma:2016phn}, can be easily evaded.
For instance, if we take $M_1\sim M_2\sim 3\times 10^{13}$ GeV with $\tan\beta={\cal O}(1)$, the neutrino mass constraint on the RPV parameter is no stronger than $\epsilon \lesssim 1$.

\subsection{Constraints on Other Relevant Operators}

Another way of encoding gravitino couplings to matter is by using the equivalence theorem \cite{equivalence} and using the Goldstino couplings, which are present in particular in the soft terms in the low-energy effective supersymmetric Lagrangian. Some of them correspond to non-universal couplings of the Goldstino to matter \cite{nonuniversal}, which are not related to the usual low-energy theorems. 
Let us denote in what follows the supersymmetry breaking spurion superfield by
\begin{equation}
X = x + \sqrt{2} \theta \psi + \theta^2 F \ , \label{oo1} 
\end{equation}
where $\psi$ is the Goldstino, $x$ its scalar partner and $F = \sqrt{3} m_{3/2} M_P$ is the supersymmetry breaking scale. Then operators containing soft terms and Goldstino couplings to Standard Model particles describe also through the equivalence theorem, the gravitino couplings to Standard Model particles. However, since the equivalence theorem is valid only for momenta well above the gravitino mass, these Goldstino couplings can only be used for high-energy processes and not for gravitino decays, which still have to be computed from the original gravitino/supercurrent interactions.

The relevant non-vanishing operators for our discussion are:

$\bullet$  The soft term associated with $\mu'$:
\begin{equation}
\frac{B_{\mu'}}{F} \int d^2 \theta\, X L H_u  \rightarrow  B_{\mu'} \tilde l h_u \ . \label{oo2}
\end{equation}
This operator generates mixing between a slepton and a Higgs, 
and can be compared with the mixing between leptons and Higgsinos.
This operator would not dominate the gravitino decay rate so long as
$B_{\mu'}/{\widetilde m}^2 < \epsilon$. If we write $B_{\mu'} = B' \mu'$,
this puts a constraint on $B' < {\widetilde m}$ having assumed that 
$\mu \sim {\widetilde m}$. In principle there is another coupling
proportional to $\mu'$ between the gravitino, leptons and the 
scalars associated with $H_u$. However, for on-shell gravitinos
(as they must be in gravitino decay), $ \gamma^\mu \psi_\mu = 0$,
causing this vertex to vanish.

$\bullet$ The gravitino coupling related to the soft term associated with $\mu$:
\begin{equation}
\frac{B_{\mu}}{F} \int d^2 \theta\, X H_u H_d \rightarrow 
\ {B\mu} \ h_u h_d   \ , \label{oo7}
\end{equation}
also vanishes for an on-shell gravitino, as does an additional 
operator proportional to $\mu$.

$\bullet$ The dimension-four operator:
\begin{equation}
\frac{c_4}{M_P} \int d^2 \theta\, (L H_u) (H_u H_d)  \rightarrow 
\ c_4\frac{\mu v^2}{M_P} \ \tilde l \, h \, , \label{oo4}
\end{equation}
where the operator is assumed to be generated at the Planck scale. This operator induces a mixing between the sleptons and the Higgs. Assuming $\mu \sim m_{\tilde l} \sim {\widetilde m} = 10^{-4} M_{P}$, one obtains the estimate
\begin{equation}
c_4 \frac{\mu v^2}{{\widetilde m}^2 M_P} \lesssim \frac{\mu'}{\mu}\,,
\end{equation}
where $\mu'$ is assumed to generate a Higgsino-neutrino mixing.
Due to the suppression from the electroweak vacuum expectation value and assuming $\mu\sim {\widetilde m}$ there is no 
meaningful constraint on $c_4$.

$\bullet$ A Giudice-Masiero-like contribution to $\mu'$ and $B_{\mu'}$
is possible if the following term is added to the K\"ahler potential:
\begin{equation}
c_{GM} ( L H_u + h.c.) \subset K
\end{equation}
leading to the shift $\mu' \rightarrow \mu' +c_{GM} m_{3/2} $
and a shift in $B_{\mu'} = B' \mu' \rightarrow  B' \mu' + c_{GM} m_{3/2}^2$.
In section \ref{OC}, we will derive a limit on $\mu'$ of order 
$\mu' \lesssim 10^{-5}$ GeV for $m_{3/2} \sim$ EeV, and this can be translated into a limit on $c_{GM} < \mu'/m_{3/2} \lesssim 10^{-14}$. 
The shift in $B'\mu'$ gives a weaker limit (again from Higgs slepton mixing), $c_{GM} < \mu' {\widetilde m}/m_{3/2}^2 \lesssim 10^{-9}$.

In the rest of the paper we will assume that the main contribution to gravitino decays comes from the bilinear $\mu'$ term and therefore that the effect of all other operators like the ones above satisfy the constraints which render them sub-dominant. 

Finally, a possible origin for a small $\mu'$ term is to assume minimal flavor violation, which can generate RPV terms with coefficients that are proportional to Yukawa couplings~\cite{Csaki:2011ge}. Even though the holomorphic spurions do not carry lepton number, a bilinear $LH_u$ term can be generated after supersymmetry breaking. 
A large suppression can then be obtained if the neutrino Yukawa coupling $y_\nu \ll 1$. A complete study of this possible origin is beyond the scope of this work.

\subsection{Limits from Lepton Number Violation}

Before concluding this section,
we note that in weak scale supersymmetric models, it is possible to derive a relatively strong limit on $\mu'$ \cite{cdeo}.
The presence of an $LH_u$ mixing term,
will induce one-to-two processes involving a Higgsino, lepton, and a gauge boson.  The thermally averaged rate at a temperature, $T$ for these lepton number violating interactions is given by
\beq
\Gamma_{1\rightarrow 2} = \frac{g^2 \theta^2 T \pi}{192 \zeta(3)} \simeq 
0.014 g^2 \frac{\mu'^2}{m_f^2}T \, ,
\label{12rate}
\eeq
where $g$ is a gauge coupling, and $\theta \simeq \mu'^2/m_f^2$ is the mixing
angle induced by $\mu'$
for a fermion with mass $m_f$. Comparing the interaction rate (\ref{12rate}) with the Hubble rate, $H \simeq  \sqrt{\pi^2 N/90}~ T^2/M_P$, where $N$ is the number of relativistic degrees of freedom at $T$, gives us the condition
\beq
\mu'^2 < 56 \sqrt{N} \frac{T}{M_P}m_f^2 \, .
\label{m'limit}
\eeq
By insisting that any lepton number violating rate involving $\mu'$ remains out-of-equilibrium while sphaleron interactions are in equilibrium, i.e., between the weak scale and $\sim 10^{12}$ GeV (where the latter is determined
by comparing the sphaleron rate $\sim \alpha_W^4 T$ to the Hubble rate), the limit (\ref{m'limit}) is strongest for $m_f \sim T$, where $T$ is of order the weak scale.  For weak scale supersymmetry, the fermion can be either a lepton or Higgsino,  $N = 915/4$
and at $T \sim 100$ GeV, 
one obtains the limit \cite{cdeo}
\beq
\mu' < 2 \times 10^{-5} {\rm GeV} \, .
\label{Eq:limit2}
\eeq
For 
weak scale supersymmetry this limit translates to $\epsilon \lesssim 10^{-7}$. This is stronger than the limit
from neutrino masses in weak scale supersymmetry models \cite{Barbier:2004ez,isy}.

In the case of high scale supersymmetry, while
the Higgsino cannot be part of the thermal bath, it can still 
mediate lepton number violating interactions, but 
the limit on $\mu'$ is significantly weaker.
For example, the process $H H \leftrightarrow L L$
will involve two insertions and is suppressed by the
supersymmetry breaking scale.
The rate can be estimated as
\beq
\Gamma_{2\rightarrow 2} \simeq 10^{-2} g^4 \frac{\mu'^4}{\mu^4 {\widetilde m}^2}T^3 \, ,
\eeq
where $\tilde m \sim \mu$ is the gaugino mass.
Setting $\Gamma_{2\rightarrow 2} < H$ gives us
\beq
\mu'^4 \lesssim 200 \sqrt{N}\,\frac{\mu^4 {\widetilde m}^2}{T M_P}\,,
\eeq
This limit should now be applied at the highest temperatures
at which sphalerons are in equilibrium ($T \sim 10^{12}$ GeV),
with $N = 427/4$. Thus
\beq
\mu' < 2 \times 10^{-7} \left(\frac{\mu {\widetilde m}^{1/2}}{{\rm GeV}^{3/2}}\right)~~{\rm GeV}\,.
\label{Eq:limit1}
\eeq
The limit on $\epsilon$ then becomes 
$\epsilon < 2 \times 10^{-7} {(\tilde m/{\rm GeV})}^{1/2}$
and for ${\widetilde m} \sim 10^{14}$ GeV, we have only
$\epsilon \lesssim 1$.


\section{Gravitino Decay} 
\label{sec:gravitino_decay}

We turn now to a more detailed derivation the gravitino decay into a gauge/Higgs boson and lepton through the RPV bilinear term.
In the supergravity Lagrangian, the relevant interaction of gravitino $\psi_\mu$ to a gauge multiplet $(A_\mu, \lambda)$ and a chiral multiplet $(\phi, \chi_L)$ is given by
\begin{eqnarray}
	{\cal L} = -\frac{i}{8M_P}\bar\lambda\gamma^\mu [\gamma^\nu,\gamma^\rho] \psi_\mu F_{\nu\rho} + \left[-\frac{i}{\sqrt2 M_P}D_\mu\phi^\dagger\bar\psi_\nu\gamma^\mu\gamma^\nu\chi_L + h.c.\right] \, .
\end{eqnarray}
Calculations of the gravitino decay width
have been previously performed in several works \cite{TY,Buchmuller:2007ui,Ibarra:2007wg,Ishiwata:2008cu,Covi:2008jy,Buchmuller:2012rc,Delahaye:2013yqa,Grefe:2014bta}\footnote{Note that our notation for $\epsilon$, which parametrizes the RPV effect, differs from the notation used in some of the literature, and introduces an overall factor of $c_\beta$ that appears in the decay widths.}.

A promising signal for observing gravitino decay through the $LH_u$ term would be a monochromatic photon-neutrino pair \cite{TY,Ibarra:2007wg,Ishiwata:2008cu,carlos,Buchmuller:2012rc}. 
In this decay channel, the bino $\tilde B$ and the neutral wino $\tilde W^0$ are related to the neutrino mass eigenstate $\nu$ by the mixing matrix $U_{\tilde B \nu}=U_{15} \approx \theta_{13}$ and $U_{\tilde W^0\nu}=U_{25}\approx \theta_{23}$, respectively, 
and thus the decay width is given by
\begin{eqnarray}
	\Gamma(\psi_\mu\to \gamma\nu) &\simeq&
	\frac{m_{3/2}^3}{64\pi M_P^2} |c_W U_{\tilde B \nu}+s_W U_{\tilde W^0 \nu}|^2
	\simeq \frac{m_{3/2}^3}{64\pi M_P^2}M_Z^2\left|\frac{\mu'}{\bar\mu}\frac{M_1-M_2}{M_1M_2} s_Wc_Wc_\beta\right|^2,
\end{eqnarray}
where the neutrino mass has been neglected, and the mixing between the bino/neutral wino and neutrino are given by
\begin{equation}
	U_{\tilde B \nu} \simeq \theta_{13} \simeq -\epsilon\frac{M_Z}{M_1} s_Wc_\beta,~~~
    U_{\tilde W^0 \nu} \simeq \theta_{23} \simeq \epsilon\frac{M_Z}{M_2} c_Wc_\beta. 
\end{equation}
For high scale supersymmetry, we see that this channel carries a significant suppression of order $(M_Z/{\widetilde m})^2$ where $M_1\sim M_2\sim {\widetilde m}$.

Similarly, we can compute the partial rate for gravitino decays into $Z$ and $\nu$, whose decay width is given by
\begin{eqnarray}
	\Gamma(\psi_\mu\to Z\nu)
	&\simeq&
	\frac{m_{3/2}^3}{64\pi M_P^2}\beta_Z^2
	\left[
		|c_W U_{\tilde W^0 \nu}-s_W U_{\tilde B \nu}|^2 F_Z \right.\nonumber\\
        &&
        \left.+ \frac{8}{3}\frac{M_Z}{m_{3/2}}{\rm Re}[(c_W U_{\tilde W^0 \nu}-s_W U_{\tilde B \nu}) (s_\beta U^*_{\widetilde H_u^0 \nu}+c_\beta U^*_{\widetilde H_d^0 \nu})] J_Z
	\right.\nonumber\\
	&&
	\left.
		+ \frac{1}{6}|s_\beta U_{\widetilde H_u^0 \nu} + c_\beta U_{\widetilde H_d^0 \nu}|^2 H_Z
	\right],
    \label{znu}
\end{eqnarray}
where
\begin{eqnarray}
	\beta_X &=& 1-\frac{M_X^2}{m_{3/2}^2},\\
	F_X &=& 1 + \frac{2}{3} \frac{M_X^2}{m_{3/2}^2} + \frac{1}{3}\frac{M_X^4}{m_{3/2}^4},\\
	J_X &=& 1 + \frac{1}{2}\frac{M_X^2}{m_{3/2}^2},\\
	H_X &=& 1 + 10\frac{M_X^2}{m_{3/2}^2} + \frac{M_X^4}{m_{3/2}^4}. 
\end{eqnarray}
As stated in Section~\ref{sec:inducedmixing}, the mixing angle between $\widetilde H_u^0$ and $\nu$ comes from $\theta^T\theta$, and is proportional to max[$M_Z^2/{\widetilde m}^2,M_Z^2/(\bar\mu {\widetilde m})$] which is negligible in our case \footnote{Note that $\theta$ given in Eq. (\ref{theta}) is the solution obtained by neglecting ${\cal O}(\theta^2)$, and thus it cannot be used to compute $U_{\widetilde H_u^0\nu}$.}.
Recall that the mixing between $\widetilde H_d^0$ and $\nu$ is given by $U_{\widetilde H_d^0 \nu}\simeq -\epsilon$.
While each term in the decay width is proportional to $\epsilon^2$, for $M_Z/{\widetilde m}\ll1$ 
the dominant term comes from the final term in (\ref{znu}) containing $U_{\widetilde H_d^0 \nu}$ and is the only term which does not lead to a 
suppression which is at least $M_Z^2/{\widetilde m}^2$ or $M_Z^2/({\widetilde m} m_{3/2})$ \cite{isy}. 
The source of this term is the gravitino decay into the longitudinal component of $Z$ leading to a relative enhancement over the 
terms involving the transverse components.
Thus for $M_Z/{\widetilde m}\ll1$,
we have
\begin{eqnarray}
	\Gamma(\psi_\mu\to Z\nu) &\simeq&
	\frac{\epsilon^2 c_\beta^2m_{3/2}^3}{384\pi M_P^2}.
\end{eqnarray}

As can be seen from Eq.(\ref{chargino}), there is mixing between $\tilde W^-$ and $l$, opening the
decay channel $\psi_\mu\to W^+ l^-$ with decay width 
\begin{equation}
	\Gamma(\psi_\mu\to W^+l^-)
	\simeq
	\frac{m_{3/2}^3}{32\pi M_P^2}\beta_W^2
	\left[
		|U_{\tilde W l}|^2 F_W + \frac{8}{3}\frac{M_W}{m_{3/2}}{\rm Re}[c_\beta U_{\tilde W l} U^*_{\widetilde H l}] J_W + \frac{1}{6}|c_\beta U_{\widetilde H l}|^2 H_W
	\right],\\
	\label{wl}
\end{equation}
where the mixing angles between charged winos/Higgsinos and neutrinos are given by
\begin{equation}
	U_{\tilde W l} \simeq\epsilon\frac{\sqrt{2}M_W}{M_2}c_\beta,~~~U_{\widetilde H l} \simeq -\epsilon - \epsilon\frac{2M_W^2}{M_2\bar\mu}s_\beta c_\beta.
\end{equation}
As in the decay channel discussed above, the final term in (\ref{wl}) carries only the suppression proportional to $\epsilon^2$
without the additional high scale supersymmetry suppression of $M_W^2/{\widetilde m}^2$ or $M_W^2/({\widetilde m} m_{3/2})$, and thus for $M_W/{\widetilde m}\ll1$,
we have
\begin{eqnarray}
	\Gamma(\psi_\mu\to W^+ l^-) &\simeq&
	\frac{\epsilon^2c_\beta^2m_{3/2}^3}{192\pi M_P^2}.
\end{eqnarray}

Finally, the longitudinal component of the gravitino also decays into $h\nu$ where $h$ is the lightest Higgs boson.
The decay width of this channel is given by
\begin{eqnarray}
	\Gamma(\psi_\mu\to h\nu) &\simeq& 
	\frac{m_{3/2}^3}{384\pi M_P^2}\beta_h^4 |s_\beta U_{\widetilde H_u^0 \nu} + c_\beta U_{\widetilde H_d^0 \nu}|^2,
\end{eqnarray}
where again the last term dominates bearing only the suppression proportional to $\epsilon^2$.

\begin{figure}[ht!]
\centering
\includegraphics[scale=.9]{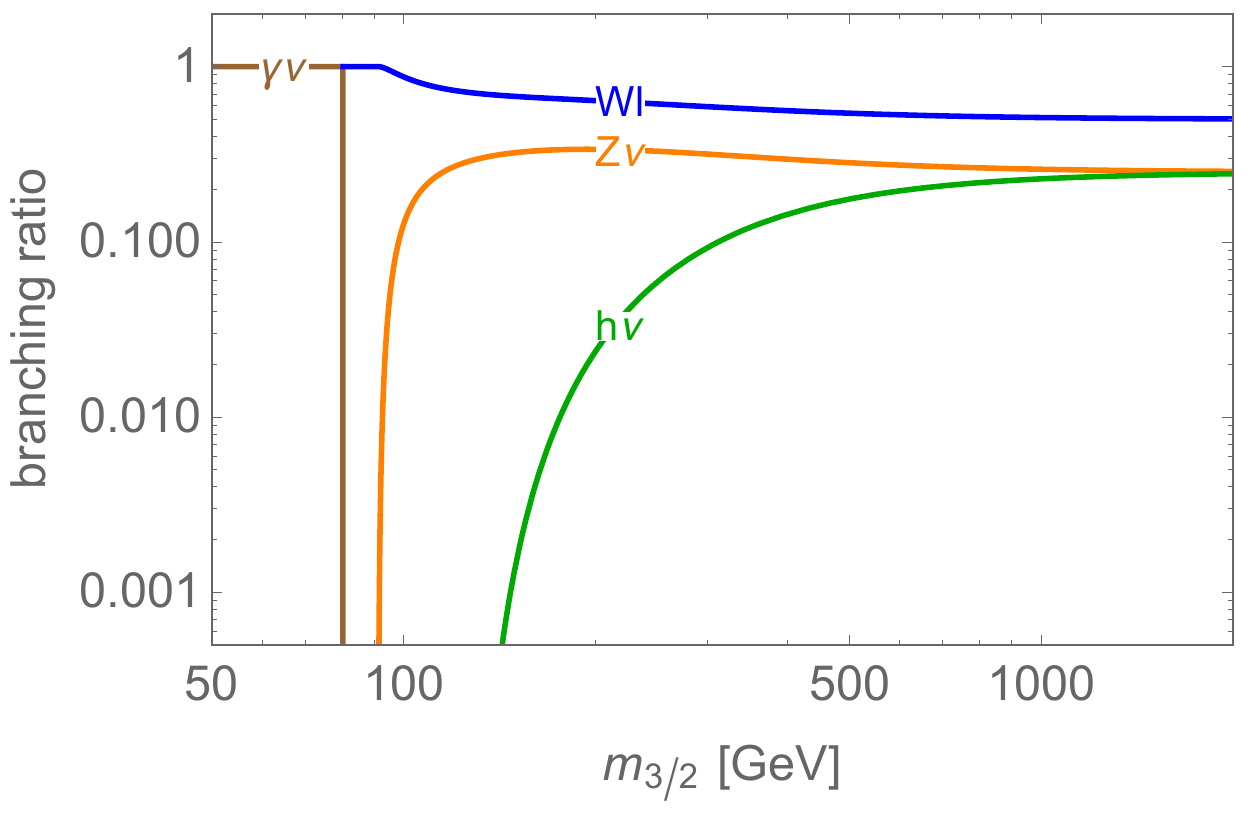}
\vskip .2in
\includegraphics[scale=.9]{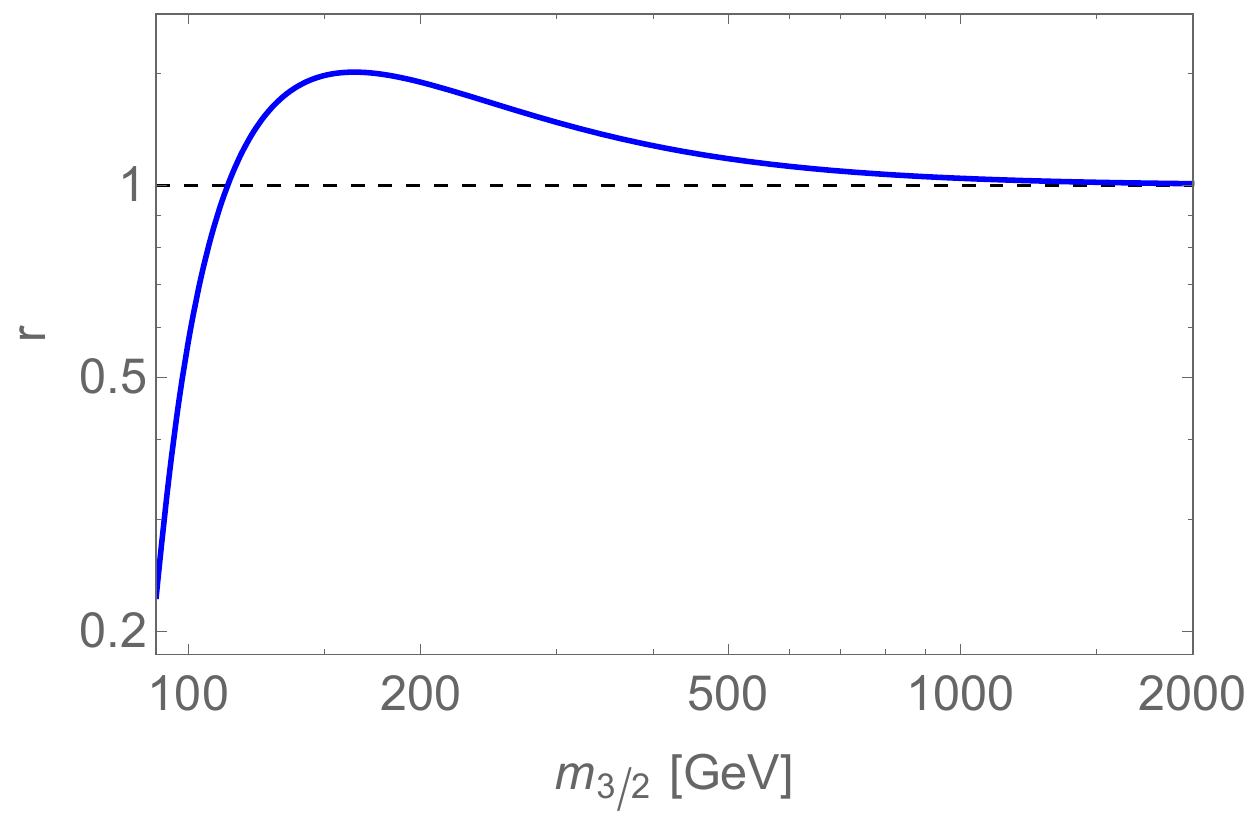}
\caption{{ Branching ratios (top) and the deviation $r$ (\ref{eq:totalratio}), from the asymptotic value for $\Gamma_{\rm tot}$ (bottom) with $M_1=M_2/2=\mu={\widetilde m}=10^{14}$ GeV.
}}
\label{Fig:decay}
\end{figure}

Figure \ref{Fig:decay} (top) shows the branching ratios of the two-body gravitino decays.
While we take $M_1=M_2/2=\mu=10^{14}$ GeV in the figure, the result is largely independent of those scales as long as ${\widetilde m} \gg {\cal O}(100)~{\rm GeV}$.
Since $M_Z/{\widetilde m}\ll 1$ in our case, $\Gamma(\psi_\mu\to \gamma\nu)$ is much smaller than $\Gamma(\psi_\mu\to W l)$, and thus the branching ratio of the $\psi_\mu\to W l$ channel dominates soon after $m_{3/2}$ becomes larger than $\sim M_W$.
For $m_{3/2}\gtrsim 1~{\rm TeV}$, the branching ratios of the decay channels $\psi_\mu\to Z\nu/Wl/h\nu$ converge to their asymptotic values with the relationship $2\Gamma(\psi_\mu\to Z\nu)=\Gamma(\psi_\mu\to Wl)=2\Gamma(\psi_\mu\to h\nu)$, as expected by the equivalence theorem.

Thus, the decay channels $\psi_\mu\to Z\nu/Wl/h\nu$ are all much larger than the $\gamma\nu$ channel for ${\widetilde m}\gg {\cal O}(100)~{\rm GeV}$, due to the enhancement
of the decay into the Higgs/Nambu-Goldstone boson (longitudinal components of the gauge bosons) which can be traced 
to the fact that the Higgsino-lepton mixings are larger than the gaugino-neutrino mixing.
In the large $m_{3/2}$ limit, each decay width is given by
\begin{eqnarray}
	\sum_i\Gamma(\psi_\mu\to Z\nu_i) &\simeq& \frac{\epsilon^2c_\beta^2m_{3/2}^3}{64\pi M_P^2},\\
    \sum_i\Gamma(\psi_\mu\to W l_i) &\simeq& \frac{\epsilon^2c_\beta^2m_{3/2}^3}{32\pi M_P^2},\\
    \sum_i\Gamma(\psi_\mu\to h\nu_i) &\simeq& \frac{\epsilon^2c_\beta^2m_{3/2}^3}{64\pi M_P^2},
\end{eqnarray}
where the charge conjugate of the final state and the number of neutrinos are incorporated\footnote{We have assumed that $\mu'$ is flavor universal.}. 
Thus the total decay width is given by
\begin{equation}
	\Gamma_{\rm tot} \simeq \frac{\epsilon^2c_\beta^2m_{3/2}^3}{16\pi M_P^2},
\end{equation}
which is indeed a good approximation for $m_{3/2}\gtrsim 1~{\rm TeV}$.
Figure \ref{Fig:decay} (bottom) shows the deviation of the total decay width from this asymptotic value with $M_1=M_2/2=\mu=10^{14}~{\rm GeV}$, which is parametrized by
\beq
	r = \Gamma_{\rm tot}/\left(\frac{\epsilon^2c_\beta^2m_{3/2}^3}{16\pi M_P^2}\right).
    \label{eq:totalratio}
\eeq
Thus, in the large $m_{3/2}$ limit, the gravitino lifetime is given by
\begin{equation}
	\tau_{3/2} \simeq 10^{28}\left(\frac{0.44\times 10^{-20}}{\epsilon c_\beta}\right)^2\left(\frac{1~{\rm EeV}}{m_{3/2}}\right)^3{\rm s}.
    \label{Eq:tau}
\end{equation}
In the next section, we derive a constraint on $\epsilon$, by ensuring that a) we have sufficient dark matter 
and b) that the decay products do not exceed observational backgrounds. 

\section{Observational Constraints}
\label{OC}

\subsection{PLANCK Constraints}

\noindent
Cosmological constraints on models with high scale supersymmetry are severe. Indeed, the only way to produce the gravitino in the early Universe if the supersymmetry breaking scale lies above the reheating temperature\footnote{To be more precise, above the maximum temperature of the thermal bath $T_{max}$ which is different from $T_{RH}$ if one considers non-instantaneous reheating \cite{gmop}.}, $T_{RH}$, is through the exchange of highly virtual sparticles with Planck-suppressed couplings,
such as t-channel processes of the type 
$G ~G~ \rightarrow ~\tilde G ~ \rightarrow ~ \psi_{\mu}~\psi_{\mu}$, with $G,\tilde G$ representing the gluon and gluino, respectively \cite{bcdm}. Because the production rate is doubly Planck-suppressed, the abundance of dark matter produced from the bath is very limited (proportional to $T_{RH}^7$ \cite{bcdm} as in Eq. (\ref{Eq:omega})), requiring a massive gravitino to compensate its low density. Moreover, it was shown in \cite{dmo,dgmo} that considering reheating processes involving inflaton decay imposes a lower bound on $T_{RH} \gtrsim 3 \times 10^{10}$ GeV implying from Eq.(\ref{Eq:omega}) a lower 
bound on the gravitino mass $m_{3/2} \gtrsim 0.2$ EeV \cite{dmo} to respect PLANCK constraints \cite{planck} on the density of cold dark matter. 

It is of interest to check this constraint in the context of models with the bilinear R-parity breaking term in  Eq.~(\ref{Wrpv}). In the context of  high scale supersymmetry, 
\beq
\mu \sim {\widetilde m} \gg \mu' ~~~ \Rightarrow ~~~\epsilon = \frac{\mu'}{\sqrt{\mu^2 + \mu'^2}} \simeq \frac{\mu'}{\mu } 
\simeq \frac{\mu'}{{\widetilde m}}\, .
\eeq
We can then rewrite Eq.(\ref{Eq:tau}):
\beq
\tau_{3/2} \simeq  10^{28}\left( \frac{{\widetilde m}}{10^{14}~\mathrm{GeV}}  \right)^2 \left( \frac{0.44~ \mathrm{keV}}{\mu'c_\beta} \right)^2 \left( \frac{1~\mathrm{EeV}}{m_{3/2}} \right)^3~\mathrm{s} \,.
\label{Eq:taubis}
\eeq
One of the interesting features in this framework is that the scale of the gravitino mass required to obtain the experimentally determined relic abundance  from Eq.~(\ref{Eq:omega}) is around the PeV-EeV scale (and higher). The decay of a particle with this mass  would provide a smoking gun signature: a monochromatic neutrino from its decay
into $Z \nu$ or $h \nu$ (Eq.~(\ref{Eq:taubis})) 
which could be observed by IceCube \cite{Icecube} or ANITA \cite{ANITA}.

Combining the relic density constraint Eq.~(\ref{Eq:omega}) 
with Eq.~(\ref{Eq:taubis}), we can eliminate the gravitino mass
and write\footnote{We have utilized non-instantaneous reheating in solving the complete set of Boltzmann equations \cite{gmop} with $T_{max}=100\times T_{RH}$}
\beq
\mu'c_\beta = 14~\mathrm{keV}\left( \frac{\Omega_{3/2} h^2}{0.11} \right)^{1/2} \left(\frac{10^{28}~\mathrm{s}}{\tau_{3/2}}  \right)^{1/2} \left(\frac{{\widetilde m}}{10^{14} \,\mathrm{GeV}} \right) \left( \frac{2.0\times 10^{10}\,\mathrm{GeV}}{T_{\mathrm{RH}}} \right)^{7/2}\, .
\label{Eq:limitmup}
\eeq
We see that while the high scale supersymmetry framework does not yield a 
strong constraint from lepton number violation ($\mu' \lesssim \mu \simeq {\widetilde m} \simeq 10^{14}~\mrm{GeV}$ from Eq. (\ref{Eq:limit1})) just requiring the lifetime to exceed the current age of the Universe ($\tau_U \simeq 4.3\times 10^{17}$ s),
would give the limit $\mu'\lesssim 20$ GeV, for $c_\beta\simeq 0.1$. However,
as we will see below, observational constraints will actually require a lifetime in excess of $10^{28}$\,s, which further restricts $\mu'  < 140 $ keV, for $c_\beta\simeq 0.1$, as given in Eq. (\ref{Eq:limitmup}). 

These limits can be contrasted with those derived in weak-scale supersymmetric models, where
$\mu' < 20~\mrm{keV}$ from the preservation of the baryon asymmetry as given in Eq.(\ref{Eq:limit2}). 
In the weak scale supersymmetry scenario, gravitinos are singly produced 
from the thermal bath and the relic abundance can be expressed as
\cite{gmop,bbb}
\beq
\Omega_{3/2} h^2 \simeq 0.11 \left( \frac{100 ~\mrm{GeV}}{m_{3/2}} \right) \left( \frac{T_{\mrm{RH}}}{2.2 \times 10^{6}~\mrm{GeV}} \right) \left( \frac{M_{1/2}}{10~\mrm{TeV}} \right)^2, 
\label{Eq:omegaclassic}
\eeq
where $M_{1/2}$ is a typical gaugino mass and we have assumed
$m_{3/2} \ll M_{1/2}$. Repeating the steps outlined above, we can
again relate $\mu'$ to the gravitino lifetime,
\beq
\mu' c_\beta \simeq 1.4 ~\mrm{keV} \left( \frac{10~\mrm{TeV}}{{\widetilde m}} \right)^{2} \left( \frac{\Omega_{3/2} h^2}{0.11} \right)^{3/2} \left( \frac{10^{28} ~\mrm{s}}{\tau_{3/2}} \right)^{1/2} \left( \frac{2.2 \times 10^{6} ~\mrm{GeV}}{T_{\mrm{RH}}} \right)^{3/2}\,,
\eeq
which is comparable to the constraint in the high-scale supersymmetry model Eq.(\ref{Eq:limitmup}) when one takes into account the adjustment in $T_{RH}$ needed to obtain the correct gravitino relic density in both limits.

As one can see, in both cases (high-scale supersymmetry and weak-scale supersymmetry) the constraints imposed on the RPV couplings from the lifetime of the gravitino (when assumed to be a dark matter candidate) are comparable or stronger than the limits imposed by the lepton number violating constraints in Eq.(\ref{Eq:limit2}) for reheating temperatures compatible with inflationary scenario.

Due to a possible signature in neutrino telescopes  such as IceCube or ANITA from the observation of ultra high energy (monochromatic) neutrinos emerging in the $Z \nu$ or $h \nu$ final states of gravitino decay, we next show that it is possible to test or set new constraints on the parameter $\mu'$ once the telescope or satellite  limits are combined with PLANCK data.

\subsection{IceCube Constraints}

We next go beyond setting the relation in Eq.(\ref{Eq:limitmup})
which sets a limit on $\mu'$ for a fixed gravitino lifetime,
and use the experimental limits from IceCube as a function of the gravitino mass and/or inflationary reheat temperature. 
Indeed, unstable gravitinos decaying into monochromatic neutrinos are severely constrained by searches from the Galactic center or the Galactic halo. The IceCube collaboration has set a lower bound on the lifetime of heavy dark matter candidates \cite{nulimit3, nulimit2, Icecubebis} (and \cite{Icecube,nulimit1} for older analyses). We can also expect gamma ray fluxes produced by $Z$-decay, and although it was shown in \cite{Kalashev:2016cre} that the gamma-ray bounds are comparable to the ones derived from neutrino fluxes, the branching fraction to gamma-rays in the model discussed here is suppressed
by $(M_Z/{\widetilde m})^2$ which is negligible.

\begin{figure}[h!]
\centering.
\includegraphics[scale=1]{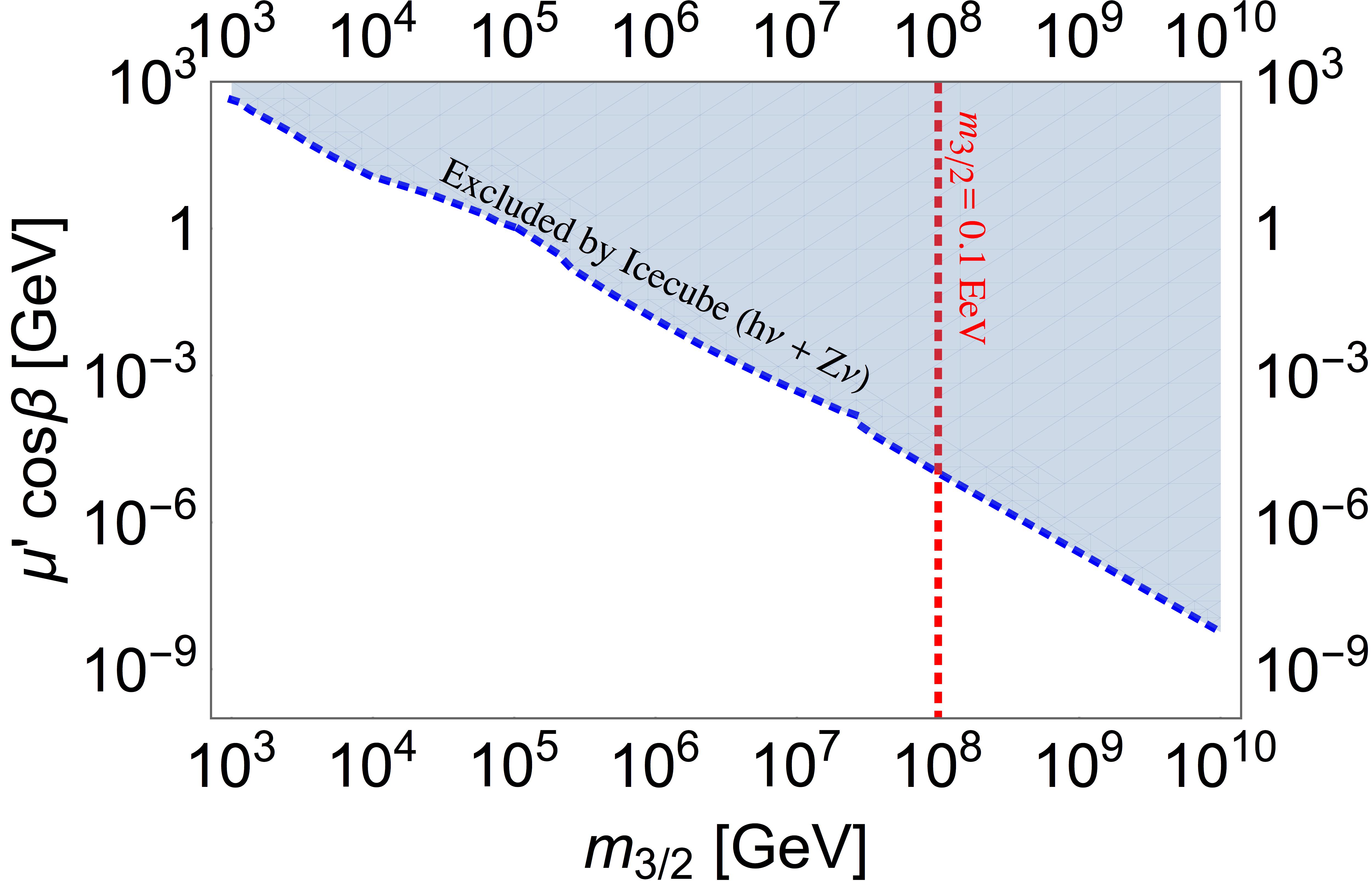}
\vskip .5in
\includegraphics[scale=1]{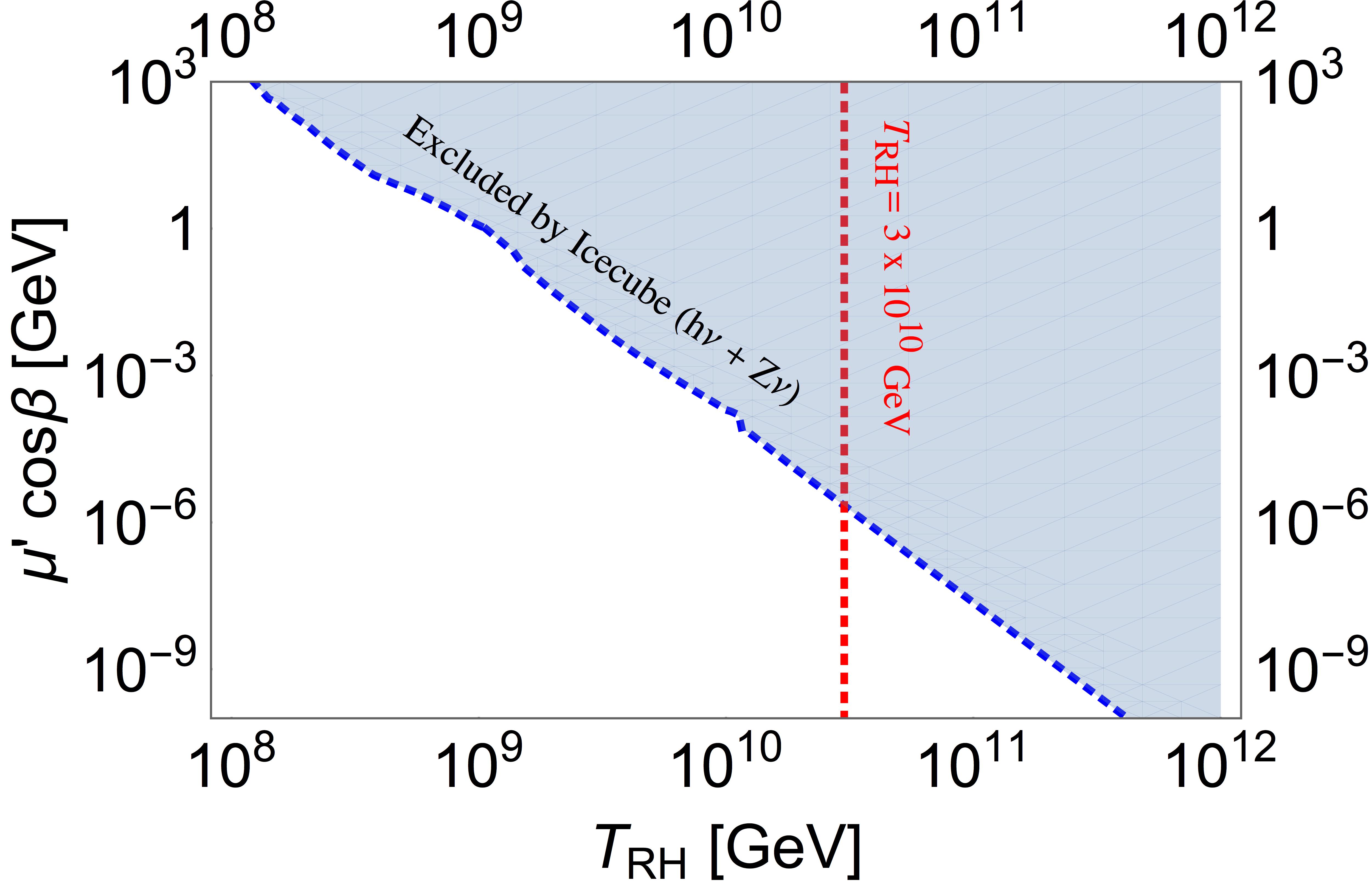}
\caption{{ Constraints from IceCube on $\mu' c_\beta$
from the $h \nu + Z\nu$ channel taking into account the relic abundance constraints from PLANCK \cite{planck} as function of the gravitino mass  (top) and as function of the reheating temperature (bottom).
}}
\label{Fig:limits}
\end{figure}

The level of interest in ultra-high energy neutrinos has been raised by the PeV events measured in the last few years by the IceCube collaboration.
IceCube recently released the combination of two of their results in \cite{nulimit3, Icecubebis}. The first analysis used 6 years  of muon-neutrino data from the northern hemisphere, while the second analysis uses 2 years of cascade data from the full sky\footnote{See also \cite{Kachelriess:2018rty} for an alternate recent analysis}. We combined both analyses ($Z \nu$ and $h \nu$ channels) with PLANCK \cite{planck} constraints to obtain limits on $\mu'$ as function of the gravitino mass and reheating temperature. 
IceCube is sensitive to energies above $\gtrsim 10^4$ GeV.
For energies of the order of the electroweak scale, we applied the limit from the Fermi satellite observation of the galactic halo \cite{fermihalo}, and the extragalactic flux \cite{fermicosmo} (see also \cite{fermicombined} for a recent combined analysis\footnote{During the completion of our work, we noticed that the MAGIC telescope released new limits on the $\nu_\tau$ cosmic flux \cite{Ahnen:2018ocv}, but these limits are currently less stringent than the ones obtained by IceCube.}).   
We present our results in Fig.~\ref{Fig:limits}.
Using Eq. (\ref{Eq:taubis}), we can set a limit on $\mu'$ as a function of $m_{3/2}$ over the mass range considered by IceCube.
Bearing in mind, that in high-scale supersymmetric models, 
we must have $m_{3/2} > 0.1$ EeV (shown by the vertical dashed line) to obtain the correct relic density \cite{dgmo}, we are confined to lower right corner in the top panel of Fig.~\ref{Fig:limits} with $\mu'\lesssim 50$ keV (for $c_\beta=0.1$) . For larger values of $\mu'$, the gravitino lifetime is too short, yielding a neutrino signal in excess of that observed by IceCube \cite{nulimit3}. Note that we have assumed a supersymmetry breaking scale of $10^{14}$ GeV, and our limit on $\mu'$  scales linearly with
${\widetilde m}$.

In the bottom panel of Fig.~\ref{Fig:limits}, we show the corresponding limit on $\mu'$ as a function of the inflationary 
reheat temperature which combines Eq.~(\ref{Eq:limitmup})
with the limit from IceCube. The vertical line at
$T_{RH}= 3 \times 10^{10}$ GeV corresponds to the lower bound on the reheating temperature if one considers inflationary-inspired models of reheating. We begin the scan at 
$T_{RH} >  5.4 \times 10^7$ GeV, corresponding to $m_{3/2} > M_Z $ extracted from Eq.~(\ref{Eq:omega}) to allow the opening of the $Z\nu$ channel. 
 Once again in order to avoid the overdensity of the Universe Eq.~(\ref{Eq:omega}), we require a massive gravitino and hence a  reheating temperature above $\sim 10^{10}$ GeV.  
 On the other hand, if we are not tied to inflationary models, there remains the possibility for $\mu' > \mathcal{O}(1)$ GeV if $T_{RH} \lesssim 10^{9}$ GeV.

\subsection{Signatures at the ANITA Experiment?}

The Antarctic Impulsive Transient Antenna (ANITA)
was designed to look for Ultra High Energy (UHE) neutrinos produced by the decay of cosmic ray products. The experiment measures radio pulses produced by the interaction of neutrinos in the ice (the Askaryan effect \cite{Askaryan:1962hbi}) and  
the balloon transporting the detector has flown three times since 2015. Recently, ANITA detected a $\sim 0.6 \pm 0.4$ EeV neutrino emerging at $27.4^o$ below the horizon \cite{ANITA1}. 
More intriguingly, an even more recent flight has observed a similar $0.56^{+0.3}_{-0.2}$ EeV
event at an angle of $35^o$ below the horizon
\cite{ANITA2}.
The measurements are consistent with the
decay of an upgoing $\tau$ generated by the interaction of an UHE $\nu_\tau$ inside the Earth.  
However, it is difficult to interpret this event as an UHE $\nu_\tau$ generated in cosmic ray fluxes because the Earth is quite opaque to such energetic $\sim $ EeV neutrinos.
Indeed, a 1 EeV neutrino has an interaction length of only 1600 kilometers water--equivalent, corresponding to an attenuation coefficient of $\sim 4 \times 10^{-6}$ for $27.4^o$ incidence angle \cite{ANITA1}.

Different explanations have been proposed, including invoking dark matter decay into a sterile neutrino \cite{Cherry:2018rxj} transforming into an active one while passing through the Earth or a
heavy 480 PeV right handed neutrino decaying into a Higgs and left-handed neutrino \cite{Anchordoqui:2018ucj}.
Both interpretations avoid the attenuation problem by the fact that sterile neutrinos have a much longer mean free path 
in water \cite{Cherry:2018rxj}. In the case of the right handed neutrino, the authors of \cite{Anchordoqui:2018ucj} claimed that the capture rate of the right handed neutrino is sufficiently strong to justify a high density of dark matter in the Earth. The probability that a dark matter particle decays not so far from the ice surface is then not negligible, and can be of the order of one decay per year as seems to be observed by ANITA.

The EeV energy measured by ANITA is particularly intriguing
as this is the mass range predicted for the gravitino in 
the high-scale supersymmetry models we are considering. 
It seems natural, therefore,  to ask whether or not an EeV gravitino could be responsible for the events observed by ANITA. 
Unfortunately, the capture rate of a gravitino by the Earth is Planck suppressed and is ridiculously low. The only possible dark matter decays which can give rise to this signal are from the local dark matter density. 
Using a local dark matter density of 0.3 GeV$\mrm{cm^{-3}}$, the radius of the Earth of 6371 kilometers, a simple computation gives, for a lifetime of $\tau_{3/2} = 1.4 \times 10^{28}$ seconds (the IceCube limit) and a gravitino mass of 0.1 EeV, the number of decaying gravitino per year $N^{\mrm{decay}}_{3/2}\simeq 1~\mrm{year} \times \frac{0.3}{m_{3/2}} \times \frac{V_{\mrm{earth}}} {\tau_{3/2}} \simeq 0.0073$ corresponding to one gravitino decaying every 137 years in the volume of the Earth\footnote{A more precise computation should be done using not the entire Earth, but only a slice corresponding to the mean free path of a 0.1 EeV neutrino, but is beyond the scope of the paper in the view of our result.}.
Although not completely ruled out, the observation of two events in 3 years seems to be in tension with our estimate.

\section{Conclusions}

While much of the high energy physics community would be overjoyed
with the detection of weak scale supersymmetry at the LHC,
we have no guarantee that the sparticle spectrum lies within the 
reach of the LHC. With the possible exception of the fine-tuning 
associated with the hierarchy problem, nearly all of the motivating
factors pointing to supersymmetry can be accounted for in either
non-supersymmetric or high-scale supersymmetric GUT models. 
In the latter we have argued that the gravitino is a dark matter candidate if its mass, $m_{3/2} > 0.1$ EeV. 

High scale supersymmetric models can be constructed so that the 
entire spectrum (except the gravitino) lies above the inflationary scale 
\cite{dgmo}. In this case, all of the superpartners of the quarks, leptons, gauge and Higgs fields, were never produced as part of the thermal bath
after inflation. For all intents and purposes, they were never
part of the physical universe. Needless to say, they would not
be produced in a laboratory/accelerator experiment. 
A stable gravitino is also experimentally problematic.
While being a perfectly good dark matter candidate from the 
point of view of gravity, its chance for detection 
in either direct or indirect detection experiments is null.

A possible escape from this conclusion of seclusion, is
the introduction of a small amount of $R$-parity violation.
Here, we considered the simplest case of the effects of an 
$\mu' LH_u$ bilinear term in the superpotential. While the limits from 
the preservation of the baryon asymmetry are greatly relaxed in high-scale supersymmetric models, the limits on this lepton number violating 
operator are strong. We have used 
the limits on the high-energy neutrino flux from IceCube
to constrain $\mu'$ as a function of the gravitino mass
and reheat temperature after inflation. For $m_{3/2} > 0.1$ EeV,
we found $\mu' < 50$ keV for $c_\beta=0.1$ (comparable to the limits obtained in weak
scale supersymmetric models). If the limit is saturated,
we would expect a signal of $\mathcal{O}(1)$ EeV neutrinos 
at IceCube and other neutrino experiments such as ANITA.
While it may be unlikely that the two high energy neutrino events observed by ANITA are related to gravitino dark matter,
this conclusion may need to be revisited if no other events
are observed in the next 140 (or so) years.

\section*{Acknowledgments}
We thank W. Buchmuller for discussions. 
This  work was supported by the France-US PICS no. 06482 and PICS MicroDark.
 Y.M. acknowledges partial support from the European Union Horizon 2020 research and innovation programme under the Marie Sklodowska-Curie: RISE InvisiblesPlus (grant agreement No 690575), the ITN Elusives (grant agreement No 674896), and the ERC advanced grants  
 Higgs@LHC. E.D. acknowledges partial support from the ANR Black-dS-String. The work of T.G., K.K., and  K.A.O. was supported in part by the DOE grant DE--SC0011842 at the University of Minnesota.

\begin{appendices}
\section{Goldstino contribution in the decay widths} 
\label{sec:goldstino_contribution_in_decay_widths}

In the decay widths given in the text, we have summed over all gravitino spin states, while there are two distinctive contributions, namely, spin $\pm3/2$ and $\pm1/2$ states.
We here discuss the spin $\pm 1/2$ Goldstino component, in the gravitino decays.
To see its contribution, it is convenient to decompose $\psi_\mu$ into spin $1$ and $1/2$ parts denoted by $\epsilon_\mu$ and $\psi$, respectively.
Incorporating Clebsch-Gordan coefficients, we have
\begin{eqnarray}
	 \psi^{\pm3/2}_\mu = \epsilon_\mu^\pm \psi^\pm,~~~
	 \psi^{\pm1/2}_\mu = \sqrt{\frac{1}{3}}\epsilon_\mu^\pm\psi^\mp + \sqrt{\frac{2}{3}}\epsilon_\mu^0\psi^\pm,
\end{eqnarray}
where $\epsilon_\mu^{\pm,0}$ and $\psi^\pm$ denote the spin $\pm1$ and 0 components for $\epsilon_\mu$ and $\pm1/2$ for $\psi$.

For the $\psi_\mu\to \gamma\nu$ decay channel, the corresponding interaction in the Lagrangian in momentum space may be written as
\begin{eqnarray}
	-\frac{i}{8M_P}\bar\lambda(k)\gamma^\mu[\gamma^\nu,\gamma^\rho]\psi_\mu(q)F_{\nu\rho}(p)
	&\sim&
	\frac{i}{4m_{3/2}M_P}\sqrt{\frac{2}{3}}\bar\lambda(k)\slashed{q}[\slashed{p},\slashed{A}(p)]\psi(q),
\end{eqnarray}
where we have used $\psi_\mu(q)\sim\sqrt{\frac{2}{3}}\frac{q_\mu}{m_{3/2}}\psi(q)$\footnote{It can be verified by a direct calculation that the contributions of the other polarization states vanish.}.
Due to the RPV coupling, the gaugino (in the gauge eigenstate) can be written as $\lambda\sim$(mixing angle)$\times \nu$ where $\nu$ is the neutrino mass eigenstate.
Also by using $q = p+k$ and the Dirac equation for $\nu$, we obtain $\bar\lambda(k)\slashed{q} \sim \bar\nu(k)(\slashed{p}+\slashed{k}) = \bar\nu(k)(\slashed{p}+m_\nu)$.
Moreover, we have $p^2=0$ and $p\cdot A=0$ for the photon, so $\slashed{p}[\slashed{p},\slashed{A}] = 0$.
Therefore, only the amplitude proportional to the neutrino mass can appear for the decay of the Goldstino mode in this channel.

On the other hand, this is not the case for the decays involving a massive gauge boson (or the Higgs boson).
For the massive gauge boson case, there appears a large enhancement for
the decay into a fermion and longitudinal mode.
In the same manner, we may write the relevant interaction as
\begin{equation}
	\frac{1}{\sqrt2 M_P} g A_\mu(p) \phi^* \bar\psi_\nu(q)\gamma^\mu\gamma^\nu \chi_L(k) + h.c.
	\simeq
	\frac{g\langle\phi\rangle}{\sqrt{3}m_{3/2}M_P} \bar\psi(q)\slashed{\epsilon}^r(p)\slashed{\epsilon}^s(q)\chi_L(k) + h.c.,
\end{equation}
where we have assumed $\phi$ and $\chi_L$ are the (up or down) Higgs and Higgsino fields, respectively, and the polarization tensors of a gauge field $A_\mu$ and gravitino are represented by $\epsilon^r(p)$ and $\epsilon^s(q)$ with $r,s$ labelling the polarization states.
Each squared amplitude denoted by $|{\cal M}(r,s)|^2$ then becomes
\begin{eqnarray}
	|{\cal M}(\pm,\pm)|^2,~|{\cal M}(\pm,0)|^2,~|{\cal M}(0,0)|^2 \sim \left(\frac{m_A}{M_P}\right)^2m_{3/2}^2,~~~|{\cal M}(0,\pm)|^2\sim\frac{m_{3/2}^4}{M_P^2},
\end{eqnarray}
where $g\langle\phi\rangle\sim m_A$ with $m_A$ the gauge boson mass, and we have taken the massless limit for $\chi_L$.
Thus, it turns out that the Goldstino mode, especially $\psi_\mu^{\pm1/2}\sim\sqrt{\frac{1}{3}}\epsilon_\mu^\pm\psi^\mp$, gives the dominant contribution in the decay into a gauge boson and neutrino pair, and by incorporating the mixing between the neutrino and Higgsino, we obtain
\begin{eqnarray}
	\Gamma(\psi_\mu\to Z_L\nu) \sim \frac{m_{3/2}^3}{M_P^2}|U_{\widetilde H\nu}|^2
	\sim
	\frac{m_{3/2}^3}{M_P^2} \epsilon^2,
\end{eqnarray}
where $Z_L$ denotes the longitudinal mode of the $Z$-boson, and $\widetilde H \approx \widetilde H_d$ which has a large mixing with the neutrino, as discussed in Section~\ref{sec:inducedmixing}.
Note that this enhancement also appears in the decay channel $\psi_\mu\to Wl$.
For the $\psi_\mu\to h\nu$ channel, the squared amplitude behaves as $m_h^2m_{3/2}^2/M_P^2$ and $m_{3/2}^4/M_P^2$ for the spin state $\psi_\mu\sim\sqrt{\frac{2}{3}}\epsilon_\mu^0\psi^\pm$ and $\sqrt{\frac{1}{3}}\epsilon_\mu^\pm\psi^\mp$, respectively, and thus, the latter is the dominant contribution and the resultant decay width becomes similar in size to the $\psi_\mu\to Z\nu,Wl$ channels.

\end{appendices}


\end{document}